\documentclass[10pt,a4paper]{article}
\usepackage[utf8]{inputenc}
\usepackage{amsmath}
\usepackage{amsfonts}
\usepackage{amssymb}
\usepackage{multirow}
\usepackage{pifont}
\usepackage{hyperref}
\usepackage[height=22cm,width=14cm]{geometry}

\begin{document}
\title{\bf{Baryon asymmetry in neutrino mass models with and without $\theta_{13}$}}

\author{Ng. K. Francis\footnote{e-mail: ngkf2010@gmail.com} \\ Department of Physics, Tezpur University, Tezpur-784028, Assam,  India
}
\date{}
\maketitle

\begin{abstract}
We investigate the comparative studies of cosmological baryon asymmetry in different neutrino mass models with and without $\theta_{13}$ 
by considering the three diagonal form of Dirac neutrino mass matrices: down quark (4,2), up-quark (8,4) and charged lepton (6,2).
The predictions of any models with $\theta_{13}$ are consistent in all the three stages of leptogenesis calculations and the results are better than the predictions of any models without $\theta_{13}$ which are consistent in a piecemeal manner with the observational data. For
the best model NH-IA (6,2) without $\theta_{13}$ , the predicted inflaton mass required to produce the observed baryon asymmetry is found to be $M_{\phi} \sim 3.6 \times 10^{10}$ GeV corresponding to reheating temperature $T_R \sim 4.5 \times 10^{16}$ GeV, while
for the same model with $\theta_{13}$: $M_{\phi} \sim 2.2 \times 10^{11}$ GeV, $T_R \sim 4.865 \times 10^{6}$ GeV and weak scale gravitino mass $m_{2/3} \sim 100$GeV without causing the gravitino problem. These values apply to the recent
discovery of Higgs boson of mass $\sim 125$ GeV. The relic abundance of gravitino is proportional to the reheating temperature of the thermal bath. One can have the right order of relic dark matter abundance only if the reheating temperature is bounded to below
$10^7$GeV.
\end{abstract}

\section{Introduction}
Recent measurement of a moderately large value of the third mixing angle $\theta_{13}$ by reactor
neutrino oscillation experiments around the world particularly by Daya Bay $(sin^2 \theta_{13}=0.089 \pm 0.010 (stat)\pm 0.005 (syst))$ \cite{1}, and RENO $(sin^2 \theta_{13}=0.113 \pm 0.013 (stat)\pm 0.019 (syst))$ \cite{2}, signifies an important breakthrough in establishing the standard three flavor oscillation picture of neutrinos. Thereby, will address the issues of the recent indication of non-maximal 2-3 mixing by MINOS accelerator experiment \cite{3} leading to determining the correct octant of $\theta_{23}$
and neutrino mass hierarchy. Furthermore, now, this has opened the door to search CP violation in the lepton sector, which in turn has profound implications for our understanding of the matter-antimatter asymmetry of the Universe. In fact, ascertaining the origin of the cosmological baryon asymmetry,$\eta_B=(6.5^{+0.4}_{-0.5})\times 10^{-10}$ \cite{4}, is one of the burning open issues in both particle physics as well as in cosmology. The asymmetry must have been generated during the evolution of the Universe. However, it is possible to dynamically generate such asymmetry if three conditions, i) the existence of baryon number violating interactions, ii) C and CP
violations and iii) the deviation from thermal equilibrium, are satisfied \cite{5}. There are different mechanisms of baryogenesis, but leptogenesis \cite{6} is attractive because of its simplicity and the connection to neutrino physics. Establishing a connection between the
low energy neutrino mixing parameters and high energy leptogenesis parameters has received much attention in recent years in Refs. \cite{6,7,8}. In leptogenesis, the first condition is satisfied by the Majorana nature of heavy neutrinos and the sphaleron effect in the
standard model (SM) at the high temperature \cite{8}], while the second condition is provided by their CP-violating decay. The deviation from thermal equilibrium is provided by the expansion of the Universe. Needless to say the departures from thermal equilibrium have
been very important-without them, the past history of the Universe would be irrelevant, as the present state would be merely that of a system at 2.75 K, very uninteresting indeed \cite{9}! One of the key to understanding the thermal history of the Universe is the estimation of cosmological baryon asymmetry from different neutrino mass models with the inclusion of the latest non-zero $\theta_{13}$.

Broadly the leptogenesis can be grouped into two: thermal with and without flavour effects and non-thermal. The simplest scenario, namely the standard thermal leptogenesis, requires nothing but the thermal excitation of heavy Majorana neutrinos which generate tiny neutrino masses via the seesaw mechanism \cite{10} and provides several implications for
the light neutrino mass spectrum \cite{11}. And with heavy hierarchical right-handed neutrino
spectrum, the CP-asymmetry and the mass of the lightest right-handed Majorana neutrino
are correlated. In order to have the correct order of light neutrino mass-squared
differences, there is a lower bound on the mass of the right-handed neutrino, $M_N \geq 10^9$ GeV \cite{12}, which in turn put constraints on reheating temperature after inflation to be $T_R \geq 10^9$ GeV. This will lead to an excessive gravitino production and conflicts with
the observed data. In the post-inflation era, these gravitino are produced in a thermal bath
due to annihilation or scattering processes of different standard particles. The relic
abundance of gravitino is proportional to the reheating temperature of the thermal bath.
One can have the right order of relic dark matter abundance only if the reheating
temperature is bounded to below $10^7$ GeV \cite{13}.On the other hand, big-bang
nucleosynthesis in SUSY theories also sets a severe constraint on the gravitino mass and
the reheating temperature leading to the upper bound $T_R \geq 10^7$
GeV \cite{14}. While thermal leptogenesis in SUSY SO(10) with high see-saw scale easily satisfies the lower bound, the
tension with the gravitino constraint is manifest.

The analysis done in Ref. \cite{15}, the non-thermal leptogenesis scenario in the framework
of a minimal supersymmetric SO (10) model with Type-I see-saw shows that the predicted
inflaton mass needed to produce the observed baryon asymmetry of the universe is found
to be $M_{\phi \sim} \sim 5 \times 10^{11}$ GeV for the reheating temperature
$T_R=10^6$ GeV and weak scale gravitino mass $m_{3/2}\sim 100$ GeV without causing the gravitino problem. It also claims that
even if these values are relaxed by one order of magnitude $(m_{3/2}\leq 10 TeV, \ T_R=10^7GeV)$, the result is still valid. In Ref. \cite{16} using the Closed-Time-Path approach, they performed a systematic leading order calculation of the relaxation rate of flavour
correlations of left-handed Standard Model leptons; and for flavoured Leptogenesis in the
Early Universe found the reheating temperature to be $T_R=10^7$ GeV to $10^{13}$ GeV. These
values apply to the Standard Model with a Higgs-boson mass of 125 GeV \cite{17}. The recent
discovery of a Standard Model (SM) like Higgs boson provides further support for
leptogenesis mechanism, where the asymmetry is generated by out-of-equilibrium decays
of our conjecture heavy right-handed neutrinos into a Higgs boson and a lepton. In
\cite{18} split neutrinos was introduced where there is one Dirac neutrino and two Majorana
with a slight departure from tribimaximal mixing (TBM), which explains the reactor angle $\sim \theta_{13}$, , and tied intimately to the lepton asymmetry and can explain inflation, dark matter, neutrino masses and the baryon asymmetry, which can be further constrained by the
searches of SUSY particles at the LHC, the right handed sneutrino, essentially the inflaton
component as a dark matter candidate, and from the $0 \nu \beta \beta$ experiments. In Ref. \cite{19} too a
deviation from TBM case was studied with model-independent and analyse the existing
link between low and high-energy parameters that connect to the parameters governing
leptogenesis. However, in Ref. \cite{20} exact TBM, $tan^2\theta_{12}=0.50$ was considered with
 charged-lepton and up-quark type and set $\theta_{13}$ zero, eventually their results differs from us.
We slightly modify the neutrino models in \cite{20}; consequently the inputs parameters are
different for zero $\theta_{13}$ but for non-zero $\theta_{13}$
our formalism is entirely different than the one
done in Ref. \cite{20}, besides we consider for  $tan^2\theta_{12}=0.45$
for detail analysis. Our work in this paper is consistent with the values given in Refs. \cite{15, 16, 17, 18}.

Now, the theoretical framework supporting leptogenesis from low-energy phases has
some other realistic testable predictions in view of non-zero $\theta_{13}$. So the present paper is a
modest attempt to compare the predictions of leptogenesis from low-energy CP-violating
phases in different neutrino mass matrices with and without $\theta_{13}$. The current investigation
is of two fold. The first part deals with zero reactors mixing angle in different neutrino
mass models within $\mu-\tau$ symmetry \cite{21}, while in the second part we construct a $m_{LL}$ matrix 
from fitting of $U_{PMNS}$ incorporating the non-zero third reactor angle $(\theta_{13})$ along with the
observed data and subsequently predict the baryon asymmetry of the Universe (BAU).

The detailed plan of the paper is as follows. In Section \ref{sec:2}, methodology and
classification of neutrino mass models for zero $\theta_{13}$
is presented. Section  \ref{sec:3}, gives an overview of leptogenesis. The numerical and analytic results for neutrino mass models $m_{LL}$
without and with $\theta_{13}$ are given in Sections \ref{sec:4} and  \ref{sec:5} respectively. We end with
summary and conclusions in Section \ref{sec:6} .

\section{Methodology and classification of neutrino mass models} 
\label{sec:2}
We begin with Type-I seesaw mechanism for estimation of BAU. The required left-handed light neutrino mass models $m_{LL}$ without $\theta_{13}$ are shifted to Appendix \textbf{A} . Since the texture of Yukawa matrix for Dirac neutrino is not known, we take the diagonal texture of $m_{LR}$  to be of charged lepton mass matrix (6,2), up-quark type mass matrix (8,4), or down-quark type mass matrix (4,2), as allowed by SO(10) GUT models.

In the second part of this paper, we construct $m_{LL}$ from $U_{PMNS}$ matrix with $\theta_{13}$ value.
\begin{equation}
m_{LL}=U_{PMNS}.m_{diag}.U^T_{PMNS}
\label{eq:1}
\end{equation}
where $U_{PMNS}$ is the PMNS  parameterised matrix taken from the standard Particle Data
Group (PDG) \cite{22}, and the corresponding mixing angles are:
\begin{equation}
sin^2 \theta_{13}=|U_{e3}|^2, tan^2\theta_{12}=\frac{|U_{e2}|^2}{|U_{e1}|^2}, tan^2\theta_{23}=\frac{|U_{\tau 3}|^2}{|U_{\mu3}|^2},
\label{eq:2}
\end{equation}

\begin{equation}
m_{diag} =
\begin{pmatrix}
m_1 & 0 & 0 \\
0 & \pm m_2 & 0 \\
0 & 0 & m_3
\end{pmatrix}.
\label{eq:3}
\end{equation}

A global analysis \cite{23} current best-fit data is used in the present analysis:
$$\Delta m^2_{21}=7.6\times 10^{-5}eV^2, \ \Delta M_{31}^2=2.4\times 10^{-3}eV^2,$$
$$sin^2 \theta_{12}=0.312, \ sin^2\theta_{23}=0.42, \ sin^2 \theta_{13}=0.025,$$
$$\theta_{12}=34^0\pm1^0, \ \theta_{23}=40.4^{+4.6^0}_{-1.8^0}, \ \theta_{13}=9.0^0\pm1.3^0.$$

Oscillation data are insensitive to the lowest neutrino mass. However, it can be measured
in tritium beta decay \cite{24}, neutrinoless double beta decay \cite{25} and from the contribution
of neutrinos to the energy density of the universe \cite{26}. Very recent data from the Planck
experiment have set an upper bound over the sum of all the neutrino mass eigenvalues of $\sum^3_{i=1}m_i \leq 0.23eV$
at $95\%$ C.L.\cite{27}. But, oscillations experiments are capable of
measuring the two independent mass-squared differences between the three neutrino
masses: $\Delta m^2_{21}=m_2^2-m^2_1$ and $\Delta m^2_{31}=m_3^2-m^2_1$. This two flavour oscillation approach
has been quite successful in measuring the solar and atmospheric neutrino parameters. In
future the neutrino experiments must involve probing the full three flavor effects,
including the sub-leading ones proportional to $\alpha=\Delta m^2_{21}/|\Delta m^2_{31}|$. The $\Delta m^2_{21}$ is positive as
is required to be positive by the observed energy dependence of the electron neutrino
survival probability in solar neutrinos but $\Delta m^2_{31}$
is allowed to be either positive or negative
by the present data. Hence, two patterns of neutrino masses are possible: $m_1 \textless m_2 \ll m_3$
called normal hierarchy (NH) where $\Delta m^2_{31}$
is positive and $m_3 \ll m_1 \textless m_2$
, called inverted hierarchy (IH) where $\Delta m^2_{31}$
is negative. A third possibility, where the three masses are nearly quasi-degenerate with very tiny differences
$m_1 \leq m_2 \sim m_3$, between them, also exists with two sub-cases of
 $\Delta^2_{31}$ being positive or negative.

Leptonic CP violation (LCPV) can be established if CP violating phase $\delta_{CP}$ is shown 
to differ from 0 and $180^0$. It was not possible to observe a signal for CP violation in the data so far.
Thus, $\delta_{CP}$ can have any value in the range [$-180^0, \ 180^0$]. The Majorana phase $\phi_1$ and $\phi_2$
are free parameters. In the absence of constraints on the phases $\phi_1$ and $\phi_2$, these have been given full variation between 0 to $2\pi$ and excluding these two extreme values.

\section{Leptogenesis}
\label{sec:3}
For our estimations of lepton asymmetry \cite{28}, we list here only the required equations for
computations. Interested reader can find more details in Ref. \cite{29}. According to Type-1
Seesaw mechanism \cite{30} the light left-handed Majorana neutrino mass matrix $m_{LL}$
heavy right-handed (RH) Majorana neutrinos $M_{RR}$, and the Dirac neutrino mass matrix $m_{LR}$
are related in a simple way

\begin{equation}
m_{LL}=-m_{LR}M^{-1}_{RR}m^{T}_{LR}
\label{eq:4}
\end{equation}

Where $m^{T}_{LR}$ is the transpose of Dirac neutrino mass matrix $m_{LR}$ and $M^{-1}_{RR}$ is the inverse of
$M_{RR}$. In unflavoured thermal leptogenesis, the lepton asymmetry generated
due to CP-violating out-of-equilibrium decay of the lightest of the heavy right-handed
Majorana neutrinos, is given by
\begin{equation}
\epsilon_1=\frac{\Gamma(N_R \rightarrow l_L+\phi)-\Gamma(N_R\rightarrow \bar{l}_L+\phi^{\dagger})}{\Gamma (N_R \rightarrow l_L+\phi)+\Gamma(N_R\rightarrow \bar{l}_L+\phi^{\dagger})}
\label{eq:5}
\end{equation}

where $\bar{l}_L$ is the anti-lepton of lepton $l_L$ and $\phi$ is the Higgs doublets chiral supermultiplets.
\begin{equation}
\epsilon_1=\frac{3}{16 \pi}\left[\frac{Im[(h^{\dagger}h)^2_{12}]}{(h^{\dagger}h)_{11}} \frac{M_1}{M_2} + \frac{Im[(h^{\dagger}h)^2_{13}]}{(h^{\dagger}h)_{11}} \frac{M_1}{M_3} \right]
\label{eq:6}
\end{equation}
where $h=m_{LR}/v$ is the Yukawa coupling of the Dirac neutrino mass matrix in the
diagonal basis of $M_{RR}$ and $v= 174$ GeV is the vev of the standard model. And finally the observed baryon asymmetry of the Universe \cite{31} is calculated from,
\begin{equation}
\eta^{SM}_B=\left(\frac{\eta_B}{\eta_{\gamma}}\right)^{SM}\approx0.98\times10^{-2} \times \kappa_1 \epsilon_1
\label{eq:7}
\end{equation}

The efficiency or dilution factor $\kappa_1$ describes the washout of the lepton asymmetry due
to various lepton number violating processes, which mainly depends on the effective
neutrino mass
\begin{equation}
\tilde{m}_1=\frac{(h^{\dagger}h)_{11}\nu^2}{M_1}
\label{eq:8}
\end{equation}
Where $v$ is the electroweak vev, $v=174$ GeV. For $10^{-2}eV < \tilde{m}_1 < 10^{-3}eV$, the washout
factor $\kappa_1$ can be well approximated by \cite{31}
\begin{equation}
\kappa_1(\tilde{m}_1)=0.3\left[\frac{10^{-3}}{\tilde{m}_1}\right]\left[log\frac{\tilde{m}_1}{10^{-3}}\right]^{-0.6}
\label{eq:9}
\end{equation}
We adopt a single expression for $\kappa_1$ valid only for the given range of $\tilde{m}_1$ \cite{32,33}.

In the flavoured thermal leptogenesis \cite{34}, we look for enhancement in
baryon asymmetry over the single flavour approximation and the equation for lepton
asymmetry in $N_1\rightarrow l_{\alpha}\phi$ decay where $\alpha=(e,\mu,\tau)$, becomes
\begin{equation}
  \varepsilon_{\alpha \alpha} = \frac{1}{8\pi}\frac{1}{(h^{\dagger}h)_{11}} \left( \sum_{j=2,3}Im\left[h^{*}_{\alpha_1}(h^{\dagger}h)_{1j}h_{\alpha j}\right]g(x_j)  + \sum_{j}Im\left[h^{*}_{\alpha_1}(h^{\dagger}h)_{j1}h_{\alpha j}\right]\frac{1}{(1-x_j)} \right)
  \label{eq:10}
\end{equation}

where $\displaystyle x_j=\frac{M^2_j}{M_i^2}$ and $\displaystyle g(x_j)\sim\frac{3}{2}\frac{1}{\sqrt{x_j}}$. The efficiency factor is given by $\displaystyle \kappa=\frac{m_*}{\tilde{m}_{\alpha \alpha}}$. Here $m_*=\frac{8\pi Hv^2}{M^2_1}\sim 1.1\times 10^{-3}$eV and $\displaystyle\tilde{m}_{\alpha \alpha}=\frac{h^{\dagger}_{\alpha 1}h_{\alpha 1}}{M_1}v^2$. This leads to the BAU,
\begin{equation}
\eta_{3B}=\frac{\eta_B}{\eta_{\gamma}}\sim10^{-2}\sum_{\alpha}\epsilon_{\alpha \alpha}\kappa_{\alpha}\sim10^{-2}m_*\sum_{\alpha}\frac{\epsilon_{\alpha \alpha}}{\tilde{m}_{\alpha \alpha}}
\label{eq:11}
\end{equation}

For single flavour case, the second term in $\epsilon_{\alpha \alpha}$ vanishes when summed over all flavours.
Thus
\begin{equation}
\epsilon_1\equiv \sum_{\alpha}\epsilon_{\alpha \alpha}=\frac{1}{8\pi}\frac{1}{(h^{\dagger}h)_{11}}\sum_j Im\left[(h^{\dagger}h)^2_{lj}\right]g(x_j),
\label{eq:12}
\end{equation}
this leads to baryon symmetry,
\begin{equation}
\eta_{1B}\approx 10^{-2}m_*\frac{\epsilon_1}{\tilde{m}}=10^{-2}\kappa_1\epsilon_1,
\label{eq:13}
\end{equation}
where $\epsilon_1=\sum_{\alpha}\epsilon_{\alpha \alpha}$ and $\tilde{m}=\sum_{\alpha}\tilde{m}_{\alpha \alpha}$.

In non-thermal leptogenesis \cite{35}the right-handed neutrinos $N_i$ $(i=1,2,3)$ with masses $(M_1, M_2, M_3)$
produced through the direct non-thermal decay of the inflaton $\phi$ interact only with leptons and Higgs through Yukawa couplings. In supersymmetric
models the superpotential that describes their interactions with leptons and Higgs is \cite{36}
\begin{equation}
W_1=Y_{ia}N_iL_aH_U
\label{eq:14}
\end{equation}
where $Y_{ia}$ is the matrix for the Yukawa couplings, $H_U$ is the superfield of the Higgs doublet that couples to up-type quarks and $L_\alpha (\alpha=e,\mu,\tau)$ is the superfield of the lepton doublet. Furthermore, for supersymmetric models the interaction between inflaton and
right-handed neutrinos is described by the superpotential \cite{37}
\begin{equation}
W_2=\sum^3_{i=1}\lambda_iSN_i^cN_i^c
\label{eq:15}
\end{equation}
 where $\lambda_i$ are the Yukawa couplings for this type of interaction and $S$
is a gauge singlet
chiral superfield for the inflaton. With such a superpotential the inflaton decay rate $\Gamma_{\phi}$
is given by \cite{37}
\begin{equation}
\Gamma_{\phi}=\Gamma \left(\phi\rightarrow N_iN_i\right)\approx \frac{|\lambda|^2}{4\pi}M_{\phi}
\label{eq:16}
\end{equation}

where $M_{\phi}$ is the mass of inflaton $\phi$. The reheating temperature ($T_R$) after inflation is \cite{38},
\begin{equation}
T_R=\left(\frac{45}{2\pi^2g}\right)^{1/4}\left(\Gamma_{\phi}M_p\right)^{1/2}
\label{eq:17}
\end{equation}
and the produced baryon asymmetry of the universe can be calculated by the
following relation \cite{39},
\begin{equation}
Y_B=\frac{n_B}{s}=CY_L=C\frac{3}{2}\frac{T_R}{M_{\phi}}\epsilon
\label{eq:18}
\end{equation}
where $s=7.0n_{\gamma}$, is related to $Y_B=n_B/S=8.7\times10^{-11}$ in Eq.(\ref{eq:18}). From Eq.(\ref{eq:18}) the connection between $T_R$ and $M_{\phi}$ is expressed as,
\begin{equation}
T_R=\left(\frac{2Y_B}{3C\epsilon}\right)M_{\phi}
\label{eq:19}
\end{equation}

Two more boundary conditions are: $M_{\phi}>2M_1$ and $T_R\leq0.01M_1$.  $ M_1$ and $\epsilon$ for 
all neutrino mass models are used in the calculation of theoretical bounds: $T_R^{min}<T_R\leq T_R^{max}$ and $M_{\phi}^{min}<M_{\phi}\leq M_{\phi}^{max}$. Only those models which satisfy these constraints can survive in the non-thermal leptogenesis.

\section{Numerical analysis and results without $\theta_{13}$}
\label{sec:4}
We first begin our numerical analysis for $m_{LL}$ without $\theta_{13}$ given in Appendix \textbf{A} . The predicted parameters for
$\tan^2\theta_{12}=0.45$, given in Table-\ref{tab:1} are consistent with the global best fit value. For leptogenesis computations, we employ the well-known inversion seesaw mechanism; $M_{RR}=-m^T_{LR}m^{-1}_{LL}m_{LR}$ and choose a basis $U_R$ where $M_{RR}^{diag}=U^T_RM_{RR}U_R=diag(M_1,M_2,M_3)$ with real and positive eigenvalues [\cite{40, 41}]. And the Dirac mass matrix $m_{LR}=diag(\lambda^m,\lambda^n,1)v$ in the prime basis transform to $m_{LR}\rightarrow m'_{LR}=m_{LR}U_RQ$, where
$Q$ is the complex matrix containing CP-violating Majorana phases $\phi_1$ and $\phi_2$ derived from $M_{RR}$. We then set the Wolfenstein parameter as $\lambda=0.3$, and compute for the three choices of $(m, n)$ in $m_{LR}$ as explained in Section \ref{sec:2} . 

In this primebasis the Dirac neutrino Yukawa coupling becomes $h=\frac{m'_{LR}}{v}$ enters in the expression of CP-asymmetry
in Eq.(\ref{eq:6}). The new Yukawa coupling matrix $h$ also becomes complex, and hence the term $Im(h^{\dagger}h)_{1j}$ appearing in lepton asymmetry $\epsilon_1$ gives a non-zero contribution. For $\phi_1$ and $\phi_2$ we choose some arbitrary values other than $\pi/2$ and 0. Finally the estimated BAU for both unflavoured $\eta_{3B}$ leptogenesis for $m_{LL}$ without $\theta_{13}$ are respectively tabulated in Table-\ref{tab:2} .

 As expected, we found that there is
an enhancement in BAU in the case of flavoured leptogenesis $\eta_{3B}$ compare to unflavoured
$\eta_{1B}$ . We also observe the sensitivity of BAU predictions on the choice of models with zero
all but the five models are favourable with good predictions [see Table-\ref{tab:2} ]. Streaming
lining further, by taking the various constraints into consideration,  QD-1A (6, 2) and NH-III (8, 4) are competing with each other, which can be tested
for discrimination in the next level-the non-thermal leptogenesis.

In case of non-thermal leptogenesis, the lightest right-handed Majorana neutrino mass $M_1$
and the CP asymmetry $\epsilon_1$ from Table-\ref{tab:2} , for all the neutrino mass models $m_{LL}$, are used in the computation of the bounds: $T^{min}_R<T_R\leq T^{max}_R$ and $M^{min}_{\phi}<M_{\phi}\leq M^{max}_{\phi}$
which are given in Table-\ref{tab:3} . The baryon asymmetry $Y_B=\frac{\eta_B}{s}$
is taken as input value from WMAP observational data. Certain inflationary models such as chaotic or natural
inflationary model predict the inflaton mass $M_{\phi}\sim 10^{13}$GeV and from Table-\ref{tab:3} , the neutrino mass models with $(m, n)$ which are compatible with $M_{\phi}\sim 10^{13}$ GeV, are listed as IA-(4, 2), IIB-(4, 2), III-(4, 2) and III-(6, 2) respectively. The neutrino mass models with (m, n)
should be compatible with $M_{\phi}\sim (10^{10}-10^{13})$ GeV. Again in order to avoid gravitino problem \cite{42} in supersymmetric models, one has the bound on reheating temperature, $T_R\approx(10^{6} - 10^{7})$Gev. This streamlines to allow models as IA-(4,2), IIB-(4,2) and III-(6,2).

\begin{table}[h]
\centering
\begin{tabular}{cccccc}
\hline 
Type & $\Delta m^2_{21}$ & $\Delta m^2_{21}$ & $tan^2\theta_{12}$ & $tan^2\theta_{23}$ & $sin\theta_{13}$ \\ 
 & $(10^{-5}eV^2)$ & $(10^{-3}eV^2)$ &  &  &  \\ 

\hline 
(IA) & 7.82 & 2.20 & 0.45 & 1.0 & 0.0 \\ 
(IB) & 7.62 & 2.49 & 0.45 & 1.0 & 0.0 \\ 
(IC) & 7.62 & 2.49 & 0.45 & 1.0 & 0.0 \\ 
\hline 
(IIA) & 7.91 & 2.35 & 0.45 & 1.0 & 0.0 \\ 
(IIB) & 8.40 & 2.03 & 0.45 & 1.0 & 0.0 \\ 
(IIC) & 7.53 & 2.45 & 0.45 & 1.0 & 0.0 \\ 
(III) & 7.61 & 2.42 & 0.45 & 1.0 & 0.0 \\ 
\hline 

\end{tabular} 
\caption{Predicted values of the solar and atmospheric neutrino mass-squared differences
and mixing angles for $tan^2\theta_{12}=0.45$}
\label{tab:1}
\end{table}

\begin{table}[h]
\centering
\begin{tabular}{ccccccc}
\hline 
Type & (m,n) & $M_1$ & $\epsilon_1$ & $\eta_{1B}$ & $\eta_{3B}$ & status \\ 
\hline 
(IA) & (4,2) & $5.43\times 10^{10}$ & $1.49\times 10^{-5}$ & $7.03\times 10^{-9}$ & $2.16\times 10^{-8}$ & \ding{51}
\\ 
(IA) & (6,2) & $4.51\times 10^{8}$ & $1.31\times 10^{-7}$ & $5.76\times 10^{-11}$ & $1.34\times 10^{-10}$ & \ding{51}
 \\ 
(IA) & (8,4) & $3.65\times 10^{6}$ & $1.16\times 10^{-9}$ & $5.72\times 10^{-13}$ & $1.19\times 10^{-12}$ & \ding{53}
 \\ 
\hline 
(IB) & (4,2) & $5.01\times 10^{9}$ & $2.56\times 10^{-14}$ & $7.15\times 10^{-15}$ & $1.09\times 10^{-9}$ & \ding{53} \\ 
(IB) & (6,2) & $4.05\times 10^{7}$ & $2.06\times 10^{-16}$ & $5.76\times 10^{-20}$ & $8.84\times 10^{-12}$ & \ding{53} \\ 
(IB) & (8,4) & $3.28\times 10^{5}$ & $1.68\times 10^{-18}$ & $4.67\times 10^{-22}$ & $7.16\times 10^{-14}$ & \ding{53} \\ 
\hline
(IC) & (4,2) & $5.01\times 10^{9}$ & $1.85\times 10^{-13}$ & $5.12\times 10^{-17}$ & $7.16\times 10^{-9}$ & \ding{53} \\ 
(IC) & (6,2) & $4.05\times 10^{7}$ & $1.47\times 10^{-15}$ & $3.77\times 10^{-29}$ & $5.80\times 10^{-11}$ & \ding{53} \\ 
(IC) & (8,4) & $3.28\times 10^{5}$ & $1.02\times 10^{-16}$ & $2.82\times 10^{-20}$ & $4.34\times 10^{-12}$ & \ding{53} \\
\hline 
(IIA) & (4,2) & $4.02\times 10^{10}$ & $1.12\times 10^{-12}$ & $2.49\times 10^{-15}$ & $7.90\times 10^{-11}$ & \ding{53}\\ 
(IIA) & (6,2) & $3.25\times 10^{8}$ & $9.00\times 10^{-15}$ & $2.00\times 10^{-17}$ & $6.34\times 10^{-13}$ & \ding{53} \\ 
(IIA) & (8,4) & $2.63\times 10^{6}$ & $7.53\times 10^{-17}$ & $1.67\times 10^{-19}$ & $5.35\times 10^{-15}$ & \ding{53} \\ 
\hline 
(IIB) & (4,2) & $9.76\times 10^{10}$ & $4.02\times 10^{-6}$ & $3.25\times 10^{-9}$ & $7.53\times 10^{-9}$ & \ding{53} \\ 
(IIB) & (6,2) & $8.10\times 10^{8}$ & $3.33\times 10^{-8}$ & $2.57\times 10^{-11}$ & $5.96\times 10^{-11}$ & \ding{53} \\ 
(IIB) & (8,4) & $6.56\times 10^{6}$ & $2.71\times 10^{-10}$ & $2.09\times 10^{-13}$ & $4.86\times 10^{-13}$ & \ding{53} \\
\hline 
(III) & (4,2) & $3.73\times 10^{12}$ & $3.09\times 10^{-5}$ & $8.13\times 10^{-8}$ & $1.85\times 10^{-6}$ & \ding{53} \\ 
(III) & (6,2) & $4.08\times 10^{11}$ & $3.74\times 10^{-5}$ & $7.37\times 10^{-10}$ & $1.62\times 10^{-9}$ & \ding{51} \\ 
(III) & (8,4) & $3.31\times 10^{9}$ & $3.09\times 10^{-7}$ & $6.06\times 10^{-11}$ & $1.13\times 10^{-10}$ & \ding{51} \\
\hline
\end{tabular} 
\caption{For zero $\theta_{13}$, lightest RH Majorana neutrino mass $M_1$ and values of CP
asymmetry and baryon asymmetry for QDN models (IA, IB, IC), IH models (IIA, IIB) and
NH models (III), with $tan^2\theta_{12}=0.45$, using neutrino mass matrices given in the
Appendix A. The entry (m, n) in $m_{LR}$ indicates the type of Dirac neutrino mass matrix
taken as charged lepton mass matrix (6, 2) or up quark mass matrix (8,4), or down quark
mass matrix (4,2) as explained in the text. IA (6,2) and III (8,4) appears to be the best
models.}
\label{tab:2}
\end{table}

\begin{table}[h]
\centering
\begin{tabular}{ccccccc}
\hline 
Type & (m,n) & $T^{min}_R<T_R \leq T^{max}_R$ & $M^{min}_{\phi}<M_{\phi} \leq M^{max}_{\phi}$ & status \\ 
\hline 
(IA) & (4,2) & $1.2\times 10^{6}<T_R \leq 5.4\times 10^{8}$ & $1.1\times 10^{11}<M_{\phi} \leq 4.9\times 10^{13}$ & \ding{51}\\ 
(IA) & (6,2) & $1.1\times 10^{6}<T_R \leq 4.5\times 10^{6}$ & $9.0\times 10^{8}<M_{\phi} \leq 3.6\times 10^{10}$ & \ding{51} \\ 
(IA) & (8,4) & $5.1\times 10^{5}<T_R \leq 3.6\times 10^{4}$ & $7.3\times 10^{6}<M_{\phi} \leq 9.6\times 10^{6}$ & \ding{53} \\ 
\hline 
(IB) & (4,2) & $6.0\times 10^{13}<T_R \leq 5.0\times 10^{7}$ & $1.0\times 10^{10}<M_{\phi} \leq 7.4\times 10^{3}$ & \ding{53} \\ 
(IB) & (6,2) & $6.4\times 10^{13}<T_R \leq 4.1\times 10^{5}$ & $8.1\times 10^{7}<M_{\phi} \leq 0.51\times 10^{1}$ & \ding{53} \\ 
(IB) & (8,4) & $6.4\times 10^{13}<T_R \leq 3.3\times 10^{3}$ & $6.6\times 10^{5} <M_{\phi} \leq 3.4\times 10^{-5}$ & \ding{53} \\ 
\hline
(IC) & (4,2) & $8.9\times 10^{12}<T_R \leq 5.0\times 10^{7}$ & $1.0\times 10^{10}<M_{\phi} \leq 5.7\times 10^{4}$ & \ding{53} \\ 
(IC) & (6,2) & $9.0\times 10^{12}<T_R \leq 4.1\times 10^{6}$ & $8.1\times 10^{7}<M_{\phi} \leq 0.36\times 10^{1}$ & \ding{53} \\ 
(IC) & (8,4) & $1.1\times 10^{12}<T_R \leq 3.3\times 10^{3}$ & $6.6\times 10^{6}<M_{\phi} \leq 1.8\times 10^{-2}$ & \ding{53} \\
\hline 
(IIA) & (4,2) & $1.3\times 10^{13}<T_R \leq 5.0\times 10^{8}$ & $8.0\times 10^{10}<M_{\phi} \leq 2.8\times 10^{6}$ & \ding{53}\\ 
(IIA) & (6,2) & $1.2\times 10^{13}<T_R \leq 4.1\times 10^{6}$ & $6.5\times 10^{8}<M_{\phi} \leq 1.8\times 10^{2}$ & \ding{53} \\ 
(IIA) & (8,4) & $1.1\times 10^{14}<T_R \leq 3.3\times 10^{4}$ & $5.3\times 10^{6}<M_{\phi} \leq 1.8\times 10^{-2}$ & \ding{53} \\ 
\hline 
(IIB) & (4,2) & $8.9\times 10^{6}<T_R \leq 5.0\times 10^{8}$ & $2.0\times 10^{12}<M_{\phi} \leq 7.0\times 10^{13}$ & \ding{53} \\ 
(IIB) & (6,2) & $8.0\times 10^{6}<T_R \leq 4.1\times 10^{6}$ & $1.6\times 10^{11}<M_{\phi} \leq 9.3\times 10^{9}$ & \ding{51} \\ 
(IIB) & (8,4) & $7.9\times 10^{6}<T_R \leq 3.3\times 10^{4}$ & $1.3\times 10^{9}<M_{\phi} \leq 6.3\times 10^{5}$ & \ding{53} \\
\hline 
(III) & (4,2) & $4.0\times 10^{7}<T_R \leq 3.7\times 10^{10}$ & $7.5\times 10^{11}<M_{\phi} \leq 7.0\times 10^{15}$ & \ding{53} \\ 
(III) & (6,2) & $3.6\times 10^{6}<T_R \leq 4.1\times 10^{9}$ & $8.2\times 10^{11}<M_{\phi} \leq 9.3\times 10^{14}$ & \ding{51} \\ 
(III) & (8,4) & $3.5\times 10^{6}<T_R \leq 3.3\times 10^{7}$ & $6.3\times 10^{9}<M_{\phi} \leq 6.3\times 10^{10}$ & \ding{51} \\
\hline
\end{tabular} 
\caption{Theoretical bound on reheating temperature $T_R$ and inflaton masses $M_{\phi}$ in non-thermal leptogenesis, for all neutrino mass models with $tan^2\theta_{12}=0.45$. Models which are consistent with observations are marked in the status column}
\label{tab:3}
\end{table}

However, the predictions of thermal leptogenesis Table-\ref{tab:2} and non-thermal
leptogenesis Table-\ref{tab:3} are not consistent for the given model [say QD-1A(6,2) or NH-
III(6,2)]; therefore, there is a problem with neutrino mass models without $\theta_{13}.$ In the next 
section, we study neutrino mass models with non-zero $\theta_{13}$ and check the consistency of above predictions.

\section{Numerical analysis and results with $\theta_{13}$}
\label{sec:5}

In this section, we investigate the effects of inclusion of non-zero $\theta_{13}$ \cite{1} \cite{2} on the
cosmological baryon asymmetry in neutrino mass models. Unlike in Section \ref{sec:4} analysis,
we don’t use the particular form of neutrino mass matrices, but we have constructed the lightest neutrino
mass matrix $m_{LL}$ using Eq.(\ref{eq:1}) through $U_{PMNS}$ and Eq.(\ref{eq:3}). Observational \cite{43} inputs used in $U_{PMNS}$ are: $\theta_{12}=34^0, \ \theta_{22}=45^0, \ \theta_{13}=9^0,$ $c_{12}=0.82904, \ c_{23}=0.707106, \ c_{13}=0.98769, \ s_{12}=0.55919, \ s_{23}=0.707106, \ c_{13}=0.156434$. We obtained

\begin{equation}
U_{PMNS} =
\begin{pmatrix}
0.81883 & 0.55230 & 0.156434 \\
-0.48711 & 0.52436 & 0.69840 \\
0.30370 & -0.64807 & 0.69840
\end{pmatrix}.
\label{eq:20}
\end{equation}

Using Eq.(\ref{eq:2}) this leads to $sin^2\theta_{13}=0.0244716, \ tan^2\theta_{12}=0.45495, \ tan^2\theta_{23}=1$. Then the $m_{diag}$ of Eq.(\ref{eq:3}) are obtained from the observational data \cite{23}  $(\Delta m^2_{12}=m^2_2-m_1^2=7.6\times 10^{-5}eV^2, \ \Delta m^2_{23}=m^2_2-m_3^2=2.4\times 10^{-3}eV^2)$, and calculated out for normal and inverted hierarchy patterns. The mass eigenvalues $m_i$ (i=1,2,3) can also be taken from Ref.\cite{29}. The positive and negative value of $m_2$ corresponds to Type-IA and Type-IB respectively. Once the matrix $m_{LL}$ is determined the procedure for subsequent calculations are same as in Section \ref{sec:4} .

Here, we give the result of only the best model due to inclusion of reactor mixing angle $\theta_{13}$  in prediction of baryon asymmetry, reheating temperature and Inflaton mass $(M_{\phi})$. Undoubtedly, for $tan^2\theta_{12} = 0.45$, the best model is NH-IA (6,2) with: baryon asymmetry in unflavoured thermal leptogenesis $B_{uf}=3.313\times 10^{-12}$; single flavoured  approximation $B_{1f}=8.844\times 10^{-12}$ and full flavoured $B_{3f}=2.093\times 10^{-11}$.
If we examine these values, we found that, expectedly, there is an enhancement is baryon asymmetry due to flavour effects. Similarly in non-thermal leptogenesis, we found that NH-IA is the best model and the predicted results are:

$$T^{min}_R<T_R\leq T_R^{max}(geV)=7.97\times 10^3<T_R\leq 4.486 \times10^6,$$
$$M^{min}_{\phi}<M_{\phi}\leq M_{\phi}^{max}(geV)=8.97\times 10^8<M_{\phi}\leq 2.24 \times10^{11}.$$

\section{Summary and conclusions}
\label{sec:6}
We now summarise the main points.  We have investigated the comparative studies of baryon asymmetry in different neutrino mass models (viz QDN, IH and NH) with and without $\theta_{13}$ for $\tan^2\theta_{12}$=0.45, and found that models with $\theta_{13}$ are better than models without $\theta_{13}$.  We found that the predictions of any models with zero $\theta_{13}$ are erratic or haphazard in spite of the fact that their predictions are consistent in a piecemeal manner with the observational data (see Tables 2 $\&$ 3) whereas the predictions of any models with non-zero $\theta_{13}$ are consistent throughout the calculations.  And among them, only the values of NH-IA (6,2) satisfied Davidson-Ibarra upper bound on the lightest RH neutrino CP asymmetry $|\epsilon_1|\leq 3.4 \times 10^{-7}$ and $M_1$ lies within the famous Ibarra-Davidson bound, i.e.,  $M_1 >4 \times 10^8$GeV \cite{44}.  Neutrino mass models either with or without $\theta_{13}$, Type-IA for charged lepton matrix (6,2) in normal hierarchy appears to be the best if $Y^{CMB}_B=6.1\times 10^{-10}$  is taken as the standard reference value, on the other hand if   then charged lepton matrix (5,2) is not ruled out.  We observed that unlike neutrino mass models with zero $\theta_{13}$,  where $\mu$ predominates over $e$ and $\tau$ contributions, for neutrino mass models with non-zero $\theta_{13}$, $\tau$ predominates over $e$ and $\mu$ contributions. This implies the  factor changes for neutrino mass models with and without $\theta_{13}$ . When flavour dynamics is included the lower bound on the reheated temperature is relaxed by a factor $\sim$ 3 to 10 as in Ref.\cite{45}.  We also observe enhancement effects in flavoured leptogenesis \cite{46} compared to non-flavoured leptogenesis by one order of magnitude as in Ref.\cite{47}. Such predictions may also help in determining the unknown Dirac Phase $\delta$ in lepton sector, which we have not studied in the present paper. The overall analysis shows that normal hierarchical model appears to be the most favourable choice in nature. Further enhancement from brane world cosmology \cite{48} may marginally modify the present findings, which we have kept for future work. 

\section*{Acknowledgements}

The author wishes to thank Prof. Ignatios Antoniadis of CERN, Geneva, Switzerland, for making comment on the manuscript and to Prof. M. K. Chaudhuri, the Vice-Chancellor of Tezpur University, for granting study leave with pay where part of the work was done during that period.  

\section*{Appendix A: Classification of neutrino mass models with zero $\theta_{13}$}

We list here the zeroth order left-handed Majorana neutrino mass matrices $m^0_{LL}$ \cite{49,50} with texture zeros left-handed Majorana neutrino mass matrices, $m_{LL}=m^0_{LL}+\Delta m_{LL}$, corresponding to three models of neutrinos, viz., Quasi-degenerate (QD1A, QD1B, QD1C), inverted hierarchical (IH2A, IH2B) and normal hierarchical (NH3) along with the inputs parameters  used in each model. $ m_{LL}$ which obey $\mu - \tau$ symmetry are constructed from their zeroth-order (completely degenerate) mass models $m^0_{LL}$  by adding a suitable perturbative term $\Delta m_{LL}$, having two additional free parameters. All the neutrino mass matrices given below predict $tan^2\theta_{12}=0.45$ . The values of three input parameters are fixed by the predictions on neutrino masses and mixings in \ref{tab:1} .

\newpage
\begin{table}[h]
\centering
\begin{tabular}{cccc}
\hline 
Type & $m^{diag}_{LL}$ & $m^{0}_{LL}$ & $m_{LL}=m^0_{LL}+\Delta m_{LL}$ \\ 
\hline 
QDIA & $diag(1,-1,1)m_0$ & $
\begin{pmatrix}
0 & \frac{1}{\sqrt{2}} & \frac{1}{\sqrt{2}}\\
\frac{1}{\sqrt{2}} & \frac{1}{2} & \frac{1}{2}\\
\frac{1}{\sqrt{2}} & -\frac{1}{2} & \frac{1}{2}
\end{pmatrix} m_0
$ & $
\begin{pmatrix}
\epsilon -2\eta & -a\epsilon & -a\epsilon \\
-a\epsilon & \frac{1}{2}-b\eta & \frac{1}{2}-\eta\\
-a\epsilon & \frac{1}{2}-\eta & \frac{1}{2}-b\eta
\end{pmatrix} m_0
$ \\ 
 &  &  &  \\
\multicolumn{4}{l}{Input $\epsilon=0.66115, \ \eta=0.16535, m_0=0.4 \ (for \ tan^2\theta_{12}=0.45,a=0.868,b=1.025)$} \\
 &  &  &  \\

 QDIB & $diag(1,1,1)m_0$ & $
\begin{pmatrix}
1 & 0 & 0\\
0 & 0 & 1\\
0 & 1 & 0
\end{pmatrix} m_0
$ & $
\begin{pmatrix}
\epsilon -2\eta & -a\epsilon & -a\epsilon \\
-a\epsilon & \frac{1}{2}-b\eta & \frac{1}{2}-\eta\\
-a\epsilon & \frac{1}{2}-\eta & \frac{1}{2}-b\eta
\end{pmatrix} m_0
$  \\ 
 &  &  &  \\
\multicolumn{4}{l}{Input $\epsilon=8.314\times 10^{-5}, \ \eta=0.00395, m_0=0.4eV \ (a=0.945,b=0.998)$} \\
 &  &  &  \\
  QD1C & $diag(1,1,-1)m_0$ & $
\begin{pmatrix}
1 & 0 & 0\\
0 & 0 & 1\\
0 & 1 & 0
\end{pmatrix} m_0
$ & $
\begin{pmatrix}
\epsilon -2\eta & -a\epsilon & -a\epsilon \\
-a\epsilon & -b\eta & 1-\eta \\
-a\epsilon & 1-\eta & -b\eta
\end{pmatrix} m_0
$  \\ 
 &  &  &  \\
\multicolumn{4}{l}{Input $\epsilon=8.211\times 10^{-5}, \ \eta=0.00395, m_0=0.4eV \ (a=0.945,b=0.998)$} \\
 &  &  &  \\
 \hline

IH2A & $diag(1,1,0)m_0$ & $
\begin{pmatrix}
1 & 0 & 0 \\
0 & \frac{1}{2} & \frac{1}{2} \\
0 & \frac{1}{2} & \frac{1}{2}
\end{pmatrix} m_0
$ & $
\begin{pmatrix}
\epsilon -2\eta & -\epsilon & -\epsilon \\
-\epsilon & \frac{1}{2} & \frac{1}{2}-\eta\\
-\epsilon & \frac{1}{2}-\eta & \frac{1}{2}
\end{pmatrix} m_0
$ \\ 
\multicolumn{4}{l}{Inverted Hierarchy with even CP parity in the first two eigenvalues (IIA)},\\
\multicolumn{4}{l}{ $(m_i=m_1,m_2,m_3): \ \frac{\eta}{\epsilon}=1.0, \ \eta=0.005, \ m_0=0.045eV$} \\
  &  &  &  \\ 

 IH2B & $diag(1,1,1)m_0$ & $
\begin{pmatrix}
0 & 1 & 1\\
1 & 0 & 0\\
1 & 0 & 0
\end{pmatrix} m_0
$ & $
\begin{pmatrix}
\epsilon -2\eta & -\epsilon & -\epsilon \\
-\epsilon & \frac{1}{2} & \frac{1}{2}-\eta\\
-\epsilon & \frac{1}{2}-\eta & \frac{1}{2}
\end{pmatrix} m_0
$ \\ 
 &  &  &  \\ 
\multicolumn{4}{l}{Inverted Hierarchy with odd CP parity in the first two eigenvalues (IIB)},\\
\multicolumn{4}{l}{ $(m_i=m_1,-m_2,m_3): \ \frac{\eta}{\epsilon}=1.0, \ \eta=0.6612, \ m_0=0.045eV$} \\
\hline 
 &  &  &  \\ 

 NH3 & $diag(0,0,1)m_0$ & $
\begin{pmatrix}
1 & 0 & 0\\
0 & \frac{1}{2} & -\frac{1}{2}\\
0 & -\frac{1}{2} & \frac{1}{2}
\end{pmatrix} m_0
$ & $
\begin{pmatrix}
0 & -\epsilon & -\epsilon \\
-\epsilon & 1-\epsilon & 1-\eta\\
-\epsilon & 1+\eta & 1-\epsilon
\end{pmatrix} m_0
$ \\ 
 &  &  &  \\ 
 \multicolumn{4}{l}{Inputs are: $\frac{\eta}{\epsilon}=0.0, \ \epsilon=0.146, \ m_0=0.028eV$} \\
\hline
\end{tabular} 
\end{table}

\end{document}